\documentstyle[epsf,12pt]{article}
\newcommand {\beq}{\begin{equation}}
\newcommand {\eeq}{\end{equation}}
\newcommand {\beqa}{\begin{eqnarray}}
\newcommand {\eeqa}{\end{eqnarray}}
\newcommand {\n}{\nonumber \\}

\renewcommand{\theequation}{\thesection.\arabic{equation}}
\begin{document}
\setlength{\oddsidemargin}{0cm}
\setlength{\baselineskip}{7mm}

\begin{titlepage}
 \renewcommand{\thefootnote}{\fnsymbol{footnote}}
$\mbox{ }$
\begin{flushright}
\begin{tabular}{l}
KEK-TH-899\\
Jun. 2003
\end{tabular}
\end{flushright}

~~\\
~~\\
~~\\

\vspace*{0cm}
    \begin{Large}
       \vspace{2cm}
       \begin{center}
         {Effective Actions of Matrix Models \\
on Homogeneous Spaces}     
\\
       \end{center}
    \end{Large}

  \vspace{1cm}

\begin{center}
           Takaaki I{\sc mai}$^{2)}$\footnote
           {
e-mail address : imaitakaaki@yahoo.co.jp},
           Yoshihisa K{\sc itazawa}$^{1),2)}$\footnote
           {
e-mail address : kitazawa@post.kek.jp}\\
           Yastoshi T{\sc akayama}$^{2)}$\footnote
           {
e-mail address : takaya@post.kek.jp}{\sc and}
           Dan T{\sc omino}$^{1)}$\footnote
           {
e-mail address : dan@post.kek.jp}

        $^{1)}$ {\it High Energy Accelerator Research Organization (KEK),}\\
               {\it Tsukuba, Ibaraki 305-0801, Japan} \\
        $^{2)}$ {\it Department of Particle and Nuclear Physics,}\\
                {\it The Graduate University for Advanced Studies,}\\
									 {\it Tsukuba, Ibaraki 305-0801, Japan}\\
\end{center}

\vfill

\begin{abstract}
\noindent 
We evaluate the effective actions 
of supersymmetric matrix models on fuzzy $S^2\times S^2$
up to the two loop level.
Remarkably it turns out to be a consistent
solution of IIB matrix model.
Based on the power counting and SUSY cancellation arguments,
we can identify the 't Hooft coupling and large $N$ scaling behavior
of the effective actions to all orders.
In the large $N$ limit, the quantum corrections 
survive except in 2 dimensional limits.
They are $O(N)$ and $O(N^{4\over 3})$
for 4 and 6 dimensional spaces respectively.
We argue that quantum effects single out
4 dimensionality among fuzzy
homogeneous spaces.
\end{abstract}
\vfill
\end{titlepage}
\vfil\eject

\section{Introduction}
\setcounter{equation}{0}
Although string theory promises to tame quantum
fluctuations of space-time, it also created 
deep questions of its own. 
One of such questions is 
to explain 4 dimensionality of space-time
since the fundamental dimension in string theory
is rather 10 or 11. Traditionally
invisible dimensions are assumed to be compactified 
at Planck scale. More recently, it has been
recognized that we could as well
be living on 4 dimensional branes
with large or even infinite extra dimensions.

In any case, we certainly need to derive
4 dimensional gauge theory and gravitation from 
string theory.
This problem can be compared to quark confinement
problem in QCD. Any nonperturbative formulation of
QCD must explain it. In this respect lattice gauge theory
has been recognized as such since confinement is
a natural phenomenon in the strong coupling limit.

We believe that 
matrix models are promising approach
to investigate these nonperturbative questions
in string theory \cite{BFSS}\cite{IKKT}.
Through them, string theory 
communicates with another promising idea of
quantum space-time namely non-commutative geometry
\cite{CDS}\cite{SW}.
In fact, non-commutative(NC) gauge theory is
naturally obtained in matrix models
with non-commutative backgrounds
\cite{AIIKKT}\cite{Li}.
The gauge invariant observables
in NC gauge theory are the Wilson lines
\cite{IIKK}.
They play crucial roles to elucidate the gravitational
aspects of NC gauge theory\cite{MRS}\cite{Gross}\cite{DK}.

In the context of IIB matrix model, 
this question has been addressed from 
several different methods.
We can cite branched polymer picture\cite{BP},
complex phase effects\cite{ANG} and mean-field approximations
\cite{NS}\cite{Kyoto}.
Although the results are encouraging thus far,
it is fair to say that the problem is still far from
settled.
In particular the 4 dimensional gaussian distributions
which have been obtained in the mean-field approximation
are not realistic space-time yet.

In this respect, we find fuzzy homogeneous spaces
to be more attractive. We have successfully constructed
these spaces using IIB matrix models\cite{Mathom}.
In the semiclassical limit, they reduce to
K\"{a}hler manifolds up to dimension 6.
Although we have not constructed experimentally favored 
de-Sitter space yet, they are closely related.
Locally we obtain maximally supersymmetric Yang-Mills
theory.
Hence our model 
provides nonperturbative formulation of
supersymmetric gauge theory as well
if we can ignore gravitation
\cite{Kaplan}\cite{NRS}. 

In this paper, we investigate IIB matrix model with 
(and without) Myers terms\cite{Myers}.
By introducing a Myers term, we can construct
non-commutative gauge theory on fuzzy sphere
at classical level\cite{IKTW}.
In our previous work, we have investigated quantum corrections
of matrix models on fuzzy sphere up to the two loop level\cite{fuzS2}.
By modifying the Myers term, we can construct higher
dimensional manifolds. In this paper, we investigate the simplest
of such 4 dimensional manifolds, namely $S^2\times S^2$.

We compute the effective action up to the two loop level.
Based on the power counting and SUSY cancellation arguments,
we can identify the 't Hooft coupling and large $N$ scaling behavior
of the effective actions to all orders.
In the large $N$ limit, the quantum corrections 
survive except in 2 dimensional limits.
They are generically $O(N)$ and $O(N^{4\over 3})$
for 4 and 6 dimensional spaces respectively.

These fuzzy homogeneous spaces are possible backgrounds
in IIB matrix model as well.
Although they are not classical solutions,
they may minimize the effective action at quantum level.
With these motivations,
we also compute the effective action of IIB matrix model
without Myers term around such backgrounds.
We indeed find that fuzzy $S^2\times S^2$ is 
a nontrivial solution of IIB matrix model
at two loop level.  In conjunction with our estimates of the large $N$
scaling behavior of the quatum corrections,
we argue that 4 dimensionality is singled out 
among fuzzy homogeneous spaces.

The organization of this paper is as follows.
In section 2, we investigate effective action of
the deformed IIB matrix model whose
classical solutions contain fuzzy $S^2\times S^2$.
In section 3, we investigate the effective actions
for fuzzy $S^2\times S^2$ in IIB matrix model itself.
We conclude in section 4 with discussions.
In Appendices A and B, we explain detailed
calculations of two loop effective actions
on fuzzy $S^2\times S^2$ with and without a Myers term.

\section{Effective actions in matrix models}
\setcounter{equation}{0}

NC gauge theories on compact homogeneous spaces $G/H$
can be constructed through matrix models.
For this purpose,
we may deform IIB matrix model as follows
\cite{Mathom}
\beq
S_{IIB}\rightarrow S_{IIB}+
{i\over 3} f_{\mu\nu\rho}Tr[A_{\mu},A_{\nu}]A_{\rho} ,
\label{IIBdef}
\eeq
where $f_{\mu\nu\rho}$ is the structure constant of a compact Lie group $G$.
Since there are 10 Hermitian matrices $A_{\mu}$ in IIB matrix model,
the number of the Lie generators of $G$ cannot exceed 10 in this
construction. Within such a constraint, we can realize
fuzzy K\"{a}hler manifolds up to dimension 6 such as
$S^1=SU(2)/U(1), CP^2=SU(3)/U(2)$\cite{KN} or  $CP^3=SO(5)/U(2)$\cite{ZH}
as classical solutions of matrix models.

Since these models possess the translation invariance
\beq
A_{\mu}\rightarrow A_{\mu}+c_{\mu} ,
\eeq
and also 
\beq
\psi\rightarrow \psi+\epsilon,
\eeq
we remove these zero-modes by restricting $A_{\mu}$ and $\psi$
to be traceless.

The equation of motion is
\beq
[A_{\mu},[A_{\mu},A_{\nu}]]+i f_{\mu\rho\nu}[A_{\mu},A_{\rho}]=0 .
\eeq
The nontrivial classical solutions are
\beq
A_{\alpha}^{cl}= t^{\alpha},
~~other ~A_{\mu}^{cl} =0 ,
\label{clsol}
\eeq
where $t^{\alpha}$'s satisfy the Lie algebra of $G$
or its sub-group.
We have investigated quantum corrections in supersymmetric
matrix models on fuzzy $S^2$\cite{fuzS2}.
In this paper we extend our investigations to
higher dimensional fuzzy homogeneous space.
Although we investigate the simplest of
such manifolds: $S^2\times S^2$,
it may reveal generic features
of matrix models on homogeneous spaces.

In order to obtain NC gauge theory on fuzzy $S^2\times S^2$,
we choose $G=SU(2)\times SU(2)$ with the following $f_{\mu\nu\rho}$:
\beqa
f_{\mu\nu\rho}&=&f\epsilon_{\mu\nu\rho};
~~(\mu,\nu,\rho) \in (8,9,0),\n
f_{\mu\nu\rho}&=&f\epsilon_{\mu\nu\rho}; 
~~(\mu,\nu,\rho) \in (1,2,3),\n
other~~ f_{\mu\nu\rho}'s&=&0.
\eeqa
We investigate the following classical solutions:
\beqa
A_{\mu}^{cl}&=& f j_{\mu}\otimes 1;~~ (\mu=8,9,0),\n
A_{\mu}^{cl}&=&f 1\otimes \tilde{j}_{\mu};~~ (\mu=1,2,3),\n
other~~ A_{\mu}^{cl} &=&0 ,
\label{S2S2}
\eeqa
where $j_{\mu}$ and $\tilde{j}_{\mu}$ are angular momentum operators.
We further assume that
$j_{\mu}$ and $\tilde{j}_{\mu}$ act on
the  $n$ copies of spin $l_1$ and $l_2$ representations 
respectively with $N=n(2l_1+1)(2l_2+1)$.
In this paper, we always assume that $l_i$ are large which
implies $n<<N$.
These solutions represent $n$ coincident fuzzy $S^2\times S^2$.
We refer to 
the constituent spaces as branes in this paper.
In the limit $l_1=l_2$, our solution represents a 4 dimensional space
while it reduces to a 2 dimensional space ($S^2$) 
in the $l_2=0$ limit.

The reducibility implies that there are $n^2-1$ 
linearly independent Hermitian traceless
matrices which commute with the classical solutions. 
They form the Lie algebra of $SU(n)$.
Its Cartan subalgebra represents the relative
center of mass coordinates of the branes. 
They will be called as zero-modes in what follows.

The classical action associated with this solution is
\beq
-{f^4\over 6}N(l_1(l_1+1)+l_2(l_2+1)) .
\label{clsact}
\eeq
In the 4d limit ($l_1=l_2$), it becomes
\beq
-{f^4\over 3}Nl_1(l_1+1)
\sim -{f^4\over 12 n}N^2 .
\eeq
In the 2d limit ($l_2=0$), we obtain
\beq
-{f^4\over 6}Nl_1(l_1+1)
\sim -{f^4\over 24 n^2}N^3 .
\label{2dact}
\eeq
Since it is minimized when $l_2=0$ and $n=1$,
a single fuzzy $S^2$ with $U(1)$ gauge group is classically favored.
As for the reducible representations corresponding 
to the multiple branes, the classical action does not
depend on their relative positions(zero-modes).

In our calculation of the partition function, we divide out the 
following gauge volume of $SU(N)/Z_N$ by gauge fixing
\beq
2^{{N^2+N\over 2}-1}\pi^{N-1\over 2}
{1\over \sqrt{N}}{1\over\prod_{k=1}^{N-1} k!}  ,
\eeq
which appeared as the universal factor in \cite{KNS}.
We recall that the `exact' free energy of IIB matrix model 
is as follows in this normalization\cite{MNS}
\beq
-log(\sum_{n|N}{1\over n^2}) .
\label{exact}
\eeq

Let us denote the bosonic and fermionic zero-modes as
$x_{\mu}$ and $\xi$.
With the presence of zero-modes, we integrate massive modes
first to obtain the Wilsonian effective action which
is a functional of zero-modes.
At the one loop level, we obtain\cite{BP}:
\beqa
&&{1\over 2}Trlog(P^2\delta_{\mu\nu}
-2iF_{\mu\nu}-2if_{\mu\nu\rho}P^{\rho}+
\bar{\Xi}\Gamma_{\mu}{1\over \Gamma\cdot P}\Gamma_{\nu}\Xi)\n
&&-Trlog(P^2)
-{1\over 4}Trlog\left( (P^2+{i\over 2}F_{\mu\nu}\Gamma^{\mu\nu})
({1+\Gamma_{11}\over 2})\right) ,
\label{1lpzm}
\eeqa
where $p_{\mu}=A_{\mu}^{cl}+x_{\mu}$
and
\beqa
&&[p_{\mu},X]=P_{\mu}X,\n
&&[f_{\mu\nu},X]=F_{\mu\nu}X,~~f_{\mu\nu}=i[p_{\mu}, p_{\nu}] ,\n
&&[\xi , X]=\Xi X .
\eeqa

We first estimate (\ref{1lpzm}) in the coincident
limit where the bosonic zero-modes are small.
Since the leading contributions in the large $N$ limit
come from large eigenvalues,
we expand (\ref{1lpzm}) into 
the power series of $F_{\mu\nu},P_{\mu}$ and $\Xi$.
The leading contribution is 
\beqa
&&-2Tr[{1\over P^2}P_{\alpha}{1\over P^2}P_{\alpha}]
-2iTr[{1\over P^2}[P_{\alpha},P_{\beta}]{1\over P^2}
f_{\alpha\beta\gamma}P_{\gamma}]\n
&=&2Tr{1\over P^2}
=2n^2\sum_{jp}(2j+1)(2p+1){1\over j(j+1)+p(p+1)} .
\label{cosmo}
\eeqa
In the 4d limit ($l_1=l_2$), we evaluate it as
\beq
4log(2)nN .
\label{4dslf}
\eeq
In the 2d limit ($l_2=0$), it becomes
\beq
4n^2log(N/n) .
\label{2dslf}
\eeq

In this process, we also obtain the products of
the following polynomials which contain fermionic zero-modes:
\beqa
&&TrS^m, \n 
&&S_{\mu\nu}={1\over P^2}
\bar{\Xi}\Gamma_{\mu}{1\over \Gamma\cdot P}\Gamma_{\nu}\Xi ,
\eeqa
where $m \leq 8(n-1)$.
Since these terms are less singular than (\ref{cosmo})
in the large $N$ limit, we may estimate $TrS^m\sim O(1/f^{3m})$.
After the fermion zero-mode integration, the normalization of the
bosonic zero-mode integration measure is determines as
\beq
\int d^{10(n-1)}x{1\over f^{14(n-1)}(N/n)^{3(n-1)}}.
\label{zeroin}
\eeq

In the presence of $n$ coincident branes, we have found that
(\ref{4dslf}) and (\ref{2dslf}) scales as
$n^2$ with $N/n$ being fixed. 
Since such configurations are further suppressed by the 
phase space of bosonic zero-modes,
we conclude that the branes tend to move away from each other.

If two branes are separated by a distance $x_{\mu}$ which is much larger
than their radii $l$, we can approximate $TrS_{\mu\nu}$ as
\beq
TrS_{\mu\nu}\sim {(N/n)^2\over f^3x^2}
\bar{\Xi}\Gamma_{\mu}{1\over \Gamma\cdot x}\Gamma_{\nu}\Xi .
\eeq
After integrating fermion zero-modes, we obtain the following
potential between them 
with respect to the identical bosonic zero-mode integration measure
as in (\ref{zeroin}).
\beqa
&&24log (|x |/l)+8log l ~~(2d~limit) ,\n
&&24log (|x |/l)-8log l ~~(4d~limit) .
\label{ldspot}
\eeqa
On the other hand, (\ref{cosmo}) can be estimated as 
\beq
-16 l^2{1\over x^2} .
\eeq
In the limit of $|x|>>l$, the former dominates the latter.
Therefore the bosonic zero-mode integration converges at large
distance and branes cannot move away from their neighbors much
farther than their radii.

Let us consider loosely bound branched polymer like configurations
of branes which are separated from their neighbors
by a distance $d >>l$.
The one loop level effective action corresponding to such a 
configuration can be estimated as
\beqa
&&16log(2)nl^2 + (n-1)log(f^{14}/l^{12})
+14(n-1)log(d/l)~~(4d~limit),\n
&&4nlog(2l) +(n-1)log(f^{14}l)
+14(n-1)log(d/l)~~(2d~limit) .
\label{uprbd}
\eeqa
This is the best upper bound of the one loop effective action
we can obtain so far.
We can interpret the first term in each limit 
as the one-loop self-energy of branes and the remaining terms
as their interactions.
We can trust the one loop estimation of brane interactions
as long as they are well separated.
Since the effective action is bounded by $O(n)$ quantity,
it is consistent with our argument that $n$ coincident branes 
cannot overlap each other. 
We thus argue that $U(n)$ gauge symmetry is broken down to $U(1)$
at the one loop level.

We find it likely that branes
settle into branched polymer like configurations
by barely touching each other.
\footnote{
In the case of the 3d model with two component Majorana spinor
\cite{IKTW}, 
the fermion
zero-mode integration results in the vanishing partition function.
Such an effect may suppress the formation of the branched polymer like
configurations.}
From the both limits in (\ref{uprbd}), we observe that
the one loop effective action favors the 2d space ($S^2$)
over the 4d space ($S^2\times S^2$) just like the tree action.

We move on to study two loop corrections.
Our strategy has been to obtain the Wilsonian effective action
by integrating massive modes with fixed zero-modes.
We delegate the detailed evaluation of
the two loop effective actions to Appendix A.
We first consider the 2d limit ($l_2=0$) which has been studied in
\cite{fuzS2}. The only novelty here is that we have the twice
contribution from the Myers term (diagram (c) in Figure 1).

The two loop effective action with $U(n)$ gauge group is
\beqa
F(l,0)&=&{1\over f^4}\left(n^3 (-40F_3^p(l)
+45F_5(l))
-n (-40F_3^{np}(l)+45F_5(l))\right)\n  
&\sim&-35{1\over f^4 (2l+1)}n(n^2-1)  +O({1/ N^2}).
\eeqa
where 
\beqa
&&F_3^p(l)\sim F_3^{np}(l)\sim {2.0\over 2l+1} ,\n
&&F_5 \sim {1\over (2l+1)} .
\eeqa
We summarize 
the effective action up to the two loop 
to the leading order of $1/N$ as
\beq
-{f^4\over 24n^2}N^3
+4n^2log(N/n)
-35{1\over f^4 N}n^2(n^2-1) .
\eeq

We still need to evaluate the higher order
corrections beyond the two loop.
Although we do not evaluate them explicitly in this paper,
we can estimate their magnitude to all orders
in perturbation theory based on our investigations thus far.
In Feynman amplitudes of matrix models on $S^2$, 
the momentum integrations of field theory
are replaced by finite series.
Our important observation is that
the large $N$ limit of the series
can be estimated by the power counting arguments
in field theory.
It is well known that there is no
ultraviolet divergences beyond the two loop level
in 2d gauge theory.
We may then conclude that all series
are convergent in the large $N$ limit in matrix models
beyond two loop.

From this observation,
we can estimate the $i$-th loop planar contribution
to be $O(n^{i+1}/(f^4l)^{i-1})$.
It is because a single factor of $1/l$ arises at each order
due the $6j$ symbols in the interaction vertices.
It thus appears that the $i$-th loop contribution is
$n^2O((\lambda_{CM}^2)^{i-1})$ where $\lambda_{CM}^2=4\pi n^2/(f^4N)$ is the
't Hooft coupling of the commutative $U(n)$ gauge theory.
In this way we can estimate the planar part of the effective action 
to all orders as:
\beq
-{\pi \over 6\lambda_{CM}^2}nN^2 +4n^2log(N/n)
+n^2h_2(\lambda_{CM}^2) .
\eeq
where $h_2(\lambda_{CM}^2)$ denotes a certain function of $\lambda_{CM}^2$.

From these arguments, we find the effective action
in the 2d limit is $O(N^2)$ in the large $N$ 
limit with $\lambda_{CM}^2$ being kept fixed  where
the tree action dominates the quantum corrections.
However the coincident limits may be exceptional configurations
since we have argued that $U(n)$ gauge group is broken down to $U(1)$
by the dissociation process of branes at the one loop level.

In the case of 
a single brane with $U(1)$ gauge group, 
the two loop effective
action is suppressed by another power of $1/N$:
\beqa
F(l,0)&=&-40{1\over f^4}(F_3^p(l)-F_3^{np}(l)) 
\sim -{248\over f^4N^2} .
\eeqa
It is because the two loop amplitude is convergent in our model
and the theory becomes free (ordinary $U(1)$ gauge theory with adjoint
matter) in the infrared limit. In other words, the planar and nonplanar
contributions cancel to the leading order of $1/N$.
Since the gauge coupling of NC $U(1)$ gauge theory is
$\lambda^2=8\pi/f^4N^2$, it is natural to find $O(\lambda^2 )$
quantum corrections at two loop.

We can summarize 
the effective action up to the two loop 
to the leading order of $1/N$ as
\beq
-{f^4 \over 24}N^3
+4log(N)
-{248\over f^4N^2} .
\eeq
Since we expect the same pattern in higher orders,
the structure of the effective action to all orders
would be
\beq
-{\pi \over 3\lambda^2}N +4log(N)
+h_2'(\lambda^2) .
\label{acs2u1}
\eeq
where $h_2'$ denotes another function of $\lambda^2$.

In the absence of UV and IR contributions, 
we are left with supersymmetric $U(1)$ NC gauge theory on $R^2$
in the large $N$ limit.
The absence of $O(N)$ quantum corrections in (\ref{acs2u1}) is consistent
with the vanishing quantum corrections in the flat space.
We thus argue that the effective action for a single
brane is $O(N)$ with $\lambda^2$ being fixed.
In such a large $N$ limit, the theory is again classical
as the tree action dominates
quantum corrections.

With the presence of $n$ branes, we have argued that
the branched polymer like
configurations with $U(1)$ gauge group are preferred 
at the one loop level.
By assuming such configurations, 
the effective action for $n$ branes
can be bounded from the above as before:
\beqa
&&-{\pi \over 3\lambda^2 }N +4nlog(N/n)
+nh_2'(\lambda^2)\n
&&-7(n-1)log(\lambda ) -6(n-1)log(l) 
+14(n-1)log(d/l),
\label{2dbrpl}
\eeqa
where $\lambda^2\sim n^2/f^4N^2$.
The first and second lines in (\ref{2dbrpl}) correspond to the
self-energy and interactions of branes respectively.

Since the classical action dominates
quantum corrections in (\ref{2dbrpl}),
a single brane minimizes it
with $f^4N^2$ being fixed.
Nevertheless the whole configurations with different
numbers of branes
energetically degenerate in the strong coupling limit.
In such a situation, we argue that
multi-brane configurations
become dominant because of their large
entropy.

We have pointed out that NC gauge theory on a fuzzy $S^2$ may be related
to 2d gravity\cite{fuzS2}. 
It is well know
that 2d supergravity with the central charge $c>1$ is unstable against
branched polymer formation.
Since deformed IIB matrix model on $S^2$ with $U(1)$ gauge group 
naively corresponds to $c=8$, the formation of branched polymers in the
strong coupling appears to be consistent with such a duality.

We next investigate the 4d limit with $l_1=l_2$.
Although the effective action could be quartically divergent
by power counting, it is only quadratically divergent
in this model. Since it is dominated by UV contributions,
the effective action for $U(n)$ gauge group is well approximated by
the planar contributions as
\beq
-8{n^3\over f^4}({F}_3^p(l,l)+2{F}_4^p(l,l))
\sim -89.3{n^3\over f^4} ,
\label{plnapr}
\eeq
where we have used the following numerically estimates:
\beqa
F_3^p(l,l)&=&3.24 +O(1/l),\n
F_4^p(l,l)&=&3.96+O(1/l).
\eeqa 
We can simply put $n=1$ in (\ref{plnapr}) for $U(1)$ gauge group
in contrast to the 2d limit.

The total effective action  up to the leading order of $1/N$ is
\beq
-{f^4\over 12 n}N^2+4log(2) nN-89.3{n^3\over f^4} .
\eeq
We can estimate the $i$-th loop contribution to be
$O(n^{i+1}l^2/(f^4 l^2)^{i-1}))$ since we obtain a factor of $1/l^2$ from
$6j$ symbols in the interaction vertices. Here we also assume that 
the leading contributions cancel due to SUSY.
Under the assumption, the amplitude is quadratically divergent 
which results in an over all factor of $l^2$.  
It thus appears that the $i$-th loop contribution is 
$O(nN (\lambda^2)^{i-1})$ where $\lambda^2=(4\pi )^2n^2/(f^4N)$
is the 't Hooft coupling of $U(n)$ gauge theory.
In this way we can estimate the planar part of the effective action 
to all orders as:
\beq
\Big(-{(2\pi )^2\over 3\lambda^2} +4log(2)
+h_4(\lambda^2)\Big)nN .
\label{4defac}
\eeq
where $h_4(\lambda^2)$ denotes a function of $\lambda^2$.
Therefore we find that the 4d effective action
is $O(N)$ in the large $N$ limit with $\lambda^2$ being fixed.

The effective action 
of loosely bound branched polymer like branes 
can be estimated as follows
\beqa
&&\Big(-{(2\pi )^2\over 3\lambda^2} +4log(2)
+h_4(\lambda^2)\Big)N\n
&&-7(n-1)log(\lambda ) -19(n-1)log(l) 
+14(n-1)log(d/l) ,
\label{nbrsf}
\eeqa
where $\lambda^2=(4\pi )^2n/(f^4N)$.
Here again the first and second line correspond to
the self-energy and the interactions of branes
respectively. 
We observe that
the interactions are sub-dominant 
in comparison to the self-energy since the former
is at most $O(nlog(l))$.
Thus the favored 4d configuration
is such that it minimizes the self-energy
with $f^4N$ being fixed.
Since 't Hooft coupling $\lambda$ changes with $n$,
we need to determine $h_4(\lambda^2)$ before
answering such a question.
We will estimate the strong coupling behavior
of $h_4(\lambda^2)$ in the next section.

Nevertheless in the large $N$ scaling region with $f^4N$ fixed
which is appropriate in 4d limit,
the effective action is negative and scales as $N^2$ in 2d limit .
Therefore in such a weak coupling regime, a single fuzzy $S^2$ always
dominates in this model.
The situation will be different in IIB matrix model
which will be investigated in the next section.

We can further determine the large $N$ scaling behavior of
the effective action in 6d case
$(S^2\times S^2\times S^2)$ in an analogous way.
Such a space can be obtained as a classical solution
of a matrix model
by modifying the Myers term.
The free energy with $U(n)$ gauge group is estimated as
\beq
- {a f^4} Nl^2+b n^2l^4-c { n^3l^6\over f^4 l^3}
+\cdots ,
\eeq
where $N=n(2l+1)^3$. $a,b, c$ are calculable numerical coefficients.
The $i$-th loop contribution can be estimated
by power counting arguments as
$O(n^{i+1}l^4/(f^4)^{i-1}l^{i-1})$.
We have also assumed here that the the 
leading contributions cancel due to SUSY in this model.
Therefore the loop expansion may make sense if we
fix $\lambda^2=n/f^4l$. 
In this way we can estimate the effective action 
to all orders as:
\beq
\Big(-{a\over \lambda^2} +b
+h_6(\lambda^2)\Big)n^{2}l^4 ,
\eeq
where $h_6(\lambda^2)$ denotes another  function of $\lambda^2$.
We conclude that the 6d effective action is
$O(N^{4\over 3})$ in the large $N$ limit with 
$f^4N^{1\over 3}$ being fixed.

\section{Effective action of IIB matrix model}
\setcounter{equation}{0}

In this section, we investigate the effective
action of IIB matrix model itself on fuzzy $S^2\times S^2$.
Our investigation parallels with that in the previous section.
The classical action of IIB matrix model for a configuration
in  (\ref{S2S2}) is
\beqa
&&-{1\over 4}Tr[A_{\mu},A_{\nu}][A_{\mu},A_{\nu}]\n
&=&{f^4\over 2}N\Big(l_1(l_1+1)+l_2(l_2+1)\Big) ,
\label{treeact}
\eeqa
where $N=n(2l_1+1)(2l_2+1)$.
If we fix $f$, the 4 dimensional limit
($S^2\times S^2$) with $l_1=l_2$ is energetically favored
over the two dimensional limit ($S^2$) with $l_2=0$.
We can also observe that larger gauge groups are
favored.
This preference is just opposite to the deformed IIB matrix model
with the Myers term. 
However we still need to minimize the action
(\ref{treeact}) with respect to $f$.
Since it is not stationary, 
we need to investigate higher order contributions 
to extract physical predictions.

The leading term of the one loop effective action 
in the large $N$ limit is
\beqa
&&-Tr\Big(({1\over P^2}^4)F_{\mu\nu}F_{\nu\lambda}F_{\lambda\rho}
F_{\rho\mu}\Big)
-2Tr\Big(({1\over P^2}^4)F_{\mu\nu}F_{\lambda\rho}F_{\mu\rho}
F_{\lambda\nu}\Big)\n
&&+{1\over 2}Tr\Big(({1\over P^2}^4)F_{\mu\nu}F_{\mu\nu}F_{\lambda\rho}
F_{\lambda\rho}\Big)
+{1\over 4}Tr\Big(({1\over P^2}^4)F_{\mu\nu}F_{\lambda\rho}F_{\mu\nu}
F_{\lambda\rho}\Big)\n
&=&3n^2Tr({1\over P^2})^2-6n^2 Tr{({P_1}^2)^2+({P_2}^2)^2\over (P^2)^4}
+n^2Tr({1\over P^2})^3\n
&=&3n^2\sum_{j,p} {(2j+1)(2p+1)\over 
(j(j+1)+p(p+1))^2}-
6n^2\sum_{j,p} {(2j+1)(2p+1)(j^2(j+1)^2+p^2(p+1)^2)\over 
(j(j+1)+p(p+1))^4}\n
&&+n^2\sum_{j,p} {(2j+1)(2p+1)\over 
(j(j+1)+p(p+1))^3} .
\label{series}
\eeqa

In 2d limit, it is estimated as
\beq
-3n^2\sum_j {2j+1 \over (j(j+1))^2}
+n^2\sum_j {2j+1 \over (j(j+1))^3}
\sim -2.6 n^2 .
\eeq
In 4d limit, we can estimate it as
\beq
- n^2log(N/n) .
\eeq
The one loop corrections are found to be
sub-leading since they are
much smaller than those in the preceding section.

Our remaining task is to determine
the structure of the effective actions
to all orders.
In 2d limit,
we can show
that the tree action dominates 
over quantum corrections
by repeating the same arguments in the previous section.
It is because the Myers terms are soft in the sense that
they do not alter power counting arguments.
Since we can derive the identical long range interaction
(\ref{ldspot}) as well, branes tend to form branched polymers.
With fixed $f$, tree action favors larger $n$ in contrast to
the preceding section. Hence the
multiple brane configurations
are always favored.
The difference is that $f$ is now a parameter which describes
the scale of field expectation values. 
We need to minimize the effective action with
respect to $f$ as well. 
Since the tree action is not stationary, 
we may conclude that 2d homogeneous spaces are
after all not realized in IIB matrix model.

In 4d case, we can repeat the same argument with the
preceding section to show that
the effective action is of the following form.
\beq
({(2\pi )^2\over \lambda^2} + \tilde{h}_4(\lambda^2))nN ,
\label{4defact}
\eeq
where $\lambda^2=(4\pi )^2n^2/(f^4N)$.
$\tilde{h}_4(\lambda^2)$ is a function which describes the quantum
corrections beyond one loop.
Explicit evaluations of the two loop contribution in 4d limit 
are reported in Appendix B. Our calculation shows that
\beq
\tilde{h}_4(\lambda^2)= 1.63{\lambda^2\over (2\pi )^2}+\cdots .
\eeq

Our next task is to minimize the effective action not only
with respect to $n$ but also
with $f$ or equivalently $\lambda^2$.
Due to the existence of nontrivial quantum corrections
$\tilde{h}_4(\lambda^2)$,
$\lambda^2$ could be fixed as
$\bar{\lambda}^2$ by minimizing  (\ref{4defact})
independently of $n$. 
At the two loop level, we find 
\beq
{\bar{\lambda}^2\over (2\pi )^2}=0.78 ,
\eeq
where the effective action assumes the minimum:
$2.55 nN$.
Remarkably
we find that 4d homogeneous spaces $S^2\times S^2$
with $U(n)$ gauge group
are solutions of IIB matrix model at two loop level
as they minimize the effective action.
Since the effective action is positive definite and $O(n^2)$,
we further conclude that $U(1)$ gauge group is
favored in the coincident limit.
The discovery of a consistent solution representing a realistic 
4 dimensional space-time in IIB matrix model
is the most important result of this paper.

Although the existence of 4d homogeneous spaces
in IIB matrix model has been shown only at the two loop level,
we can argue that higher order corrections do not
destabilize it. It is because
the effective action
grows in the both weak and strong coupling directions
as we will argue shortly .
Since the effective action is believed to be
bounded by (\ref{exact}), we may assume that
$O(N)$ term is  positive definite.
With such an assumption, we can further conclude that a single 4d homogeneous
space ($n$=1) minimizes the effective action
in the coincident limit.

In the presence of $n$ identical 4d branes,
we still need to investigate their zero-modes.
We find it likely 
that branched polymer like configurations 
with $U(1)$ gauge group will be favored again.  
It is because the effective action for such a configuration
can be estimated as follows 
\beq
({(2\pi )^2\over \bar{\lambda}^2} + \tilde{h}_4(\bar{\lambda}^2))N ,
\eeq
where we have neglected the interaction terms which are at most $O(nlog(l))$.
We thus find that the effective actions for different
numbers of branes degenerate in 4d limit to the leading order
of $N$.
This degeneracy could be lifted by the
next leading
$O(l\sim \sqrt{N})$ terms in the effective action.
We have thus presented a scenario in which
a single 4d homogeneous space $S^2\times
S^2$ with $U(1)$ gauge group is realized in IIB matrix model at quantum
level. Although we hope to understand the dynamics of the zero-modes
in more detail, it is beyond the scope of this paper.

In the remainder of this section, 
we argue that the homogeneous spaces are
continuously connected to
branched polymers.
At the classical level,
we have fuzzy homogeneous spaces
made of $N$ quanta.
The distance of the neighboring quanta
is $O(fl^{1\over 2})$.
If we fix the coupling of NC gauge theory $g\sim 1/f^2l$,
the distance of the neighbors is $O(1/{g}^{1\over 2})$.
If we fix $g\sim 1/f^2{l}^{1\over 2}$ as in 6d or 2d with $U(n)$
gauge group, neighboring points are separated by distances of 
$O(l^{1\over 4}/{g}^{1\over 2})$.
They are homogeneously distributed on the respective manifolds.

We can investigate the quantum effects on the distribution
of eigenvalues as follows.
Let us introduce the norm of $A$ as $|A|^2=TrAA^{\dagger}$.
We can also split $A=A^{cl}+ a$ where
$A^{cl}$ and $a$ denote the classical and quantum fields
respectively.

In a single sphere case, we can obtain the following estimates
up to the one loop level:
\beqa
&&|A^{cl}|=fl^{3\over 2},
~~|a^{j}|={1\over f j},\n
&&{1\over N}TrA_{\mu}A_{\nu}=f^2l^2\bar{\delta}_{\mu\nu}+{1\over
f^2l}log(l)\delta_{\mu\nu} ,
\label{2dfl}
\eeqa
where $\bar{\delta}_{\mu\nu}$ is projected in the sub-space in which
$S^2$  resides.

In 4d case,
\beqa
&&|A^{cl}|=fl^2n^{1\over 2},
~~|a^{j,k}|={1\over f \sqrt{j^2+k^2}},\n
&&{1\over N}TrA_{\mu}A_{\nu}=f^2l^2\bar{\delta}_{\mu\nu}+
{n\over f^2}{\delta}_{\mu\nu} 
={\lambda N^{1\over 2}}({\delta}_{\mu\nu}+{1\over
\lambda^2}\bar{\delta}_{\mu\nu}) ,
\eeqa
where $\bar{\delta}_{\mu\nu}$ assumes the value only in $S^2\times S^2$
sub-space.

In 6d case,
\beqa
&&|A^{cl}|=fl^{5\over 2}n^{1\over 2},
~~|a^{j,k,m}|={1\over f \sqrt{j^2+k^2+m}},\n
&&{1\over N}TrA_{\mu}A_{\nu}=f^2l^2\bar{\delta}_{\mu\nu}+{nl\over
f^2}{\delta}_{\mu\nu} ={\lambda N^{1\over 2}}({\delta}_{\mu\nu}+{1\over
\lambda^2}\bar{\delta}_{\mu\nu}) ,
\eeqa
where $\bar{\delta}_{\mu\nu}$ assumes the value only in 
$S^2\times S^2\times S^2$ sub-space.

A naive extrapolation of
these weak coupling expressions suggests 
the following correlation functions 
in the strong coupling limit for the both
4d and 6d cases:
\beq
{1\over N}TrA_{\mu}A_{\nu} \sim \lambda N^{1\over 2}\delta_{\mu\nu} .
\label{amuanu}
\eeq
In the strong coupling region, we can no longer hope to relate $\lambda$ to
the classical field $A^{cl}$ as in the weak coupling region.
$\lambda$ may be better defined as the parameter to describe
the gauge invariant observables in (\ref{amuanu}).

The gauge invariant correlator in (\ref{amuanu}) 
measures the distribution of quanta.
$\lambda N^{1\over 2}$ can be interpreted as their extension.
This scaling behavior with respect to $N$ reminds us
branched polymers whose fractal dimension is 4 \cite{BP}.
For branched polymers of $N$ quanta which
are separated by a distance $d$,
the corresponding quantity is $ (dN)^{1\over 2}$.
The scaling behavior in (\ref{amuanu}) is consistent with the
branched polymers whose $d\sim\lambda^2$.

The large $\lambda$ regime corresponds to the widely
separated eigenvalue distribution of $A$. In such a regime, IIB matrix model
is well approximated by the branched polymers and whose 
effective action can be estimated as
\beq
14 Nlog(\lambda^2) .
\eeq
This is our prediction for the universal strong coupling behavior
for the effective actions of IIB matrix model in homogeneous spaces.
Our arguments here equally apply to the effective actions
in the preceding section.

In 2d, the quantum fluctuation in (\ref{2dfl}) appear to be
smaller than the classical field even with large $g\sim 1/f^2l$
which is always smaller than $l$ in the large $N$ limit.
However we have argued in section 2 that branched polymers 
with large numbers of constituents will be formed
in the strong coupling limit
which is consistent with (\ref{amuanu}).

In this section we have found that
the effective action of IIB matrix model possesses
the minimum with finite gauge coupling $\lambda$ in 4d among 
the simplest fuzzy homogeneous spaces. 
The effective action favors 4 dimensionality since it is
$O(N)$ in 4d while it is $O(N^{4\over 3})$ in 6d.
On the other hand, it does not possess 2 dimensional solutions like $S^2$
since the theory is found to be classical in the large $N$ limit.
What is significant in our observation is that quantum fluctuations
naturally select 4 dimensionality. 
Furthermore homogeneous
spaces may be smoothly connected with the branched polymers
in the strong coupling limit.

\section{Conclusions and Discussions}
\setcounter{equation}{0}
In this paper, we have investigated two loop effective actions
of supersymmetric matrix models on a 4 dimensional fuzzy manifold
$S^2\times S^2$.
We find it remarkable that
fuzzy $S^2\times S^2$ turns out be 
a nontrivial solution of IIB matrix model
at two loop level.
Based on the power counting and SUSY cancellation arguments,
we have identified the 't Hooft coupling and large $N$ scaling behavior
of the effective actions to all orders.
In the large $N$ limit, the quantum corrections 
survive except in 2 dimensional limits.
They are generically $O(N)$ and $O(N^{4\over 3})$
for 4 and 6 dimensional spaces respectively.
These general arguments validate our solution 
beyond two loop level.

Although we have investigated the simplest 4 dimensional
manifold, we believe our results
capture generic feature of matrix models on homogeneous spaces.
It is because we have employed 
power counting and SUSY cancellation arguments 
which do not depend on detailed group structures
of homogeneous spaces.
Therefore we believe that 4 dimensionality
is a generic prediction of IIB matrix model
among fuzzy homogeneous spaces.

Although 4 dimensional distributions are clearly
favored in the mean field approximation, 
we believe we have made a substantial improvements
from the optimization point of view as well.
We have obtained $O(N)$ free energy while it is
$O(N^2)$ in mean field approximation.
However we still need to lower it down to
the prediction in (\ref{exact}).
We certainly hope to understand what kind of space-time
achieves such a feat.

\begin{center} \begin{large}
Acknowledgments
\end{large} \end{center}
This work is supported in part by the Grant-in-Aid for Scientific
Research from the Ministry of Education, Science and Culture of Japan.

\section*{Appendix A}
\renewcommand{\theequation}{A.\arabic{equation}}
\setcounter{equation}{0}

In this appendix, we evaluate 
the two loop effective action 
of NC gauge theory on $S^2\times S^2$.
We consider NC gauge theory with $U(1)$ gauge group in the context of
a deformed IIB matrix model with a Myers term. 
It is straightforward to generalize our results to the case of 
$U(n)$ gauge groups.

The evaluation procedure parallels to that of
NC gauge theory on $S^2$.
There are 5 diagrams to evaluate which are
illustrated in  Figure 1.
(a),(b) and (c) represent contributions from gauge fields.
(a) and (b) are of different topology while (c)
involves the Myers type interaction.
(d) involves ghost and (e) fermions respectively.

\begin{figure}[hbtp]
\epsfysize=3cm
\begin{center}
\vspace{1cm}
\epsfbox{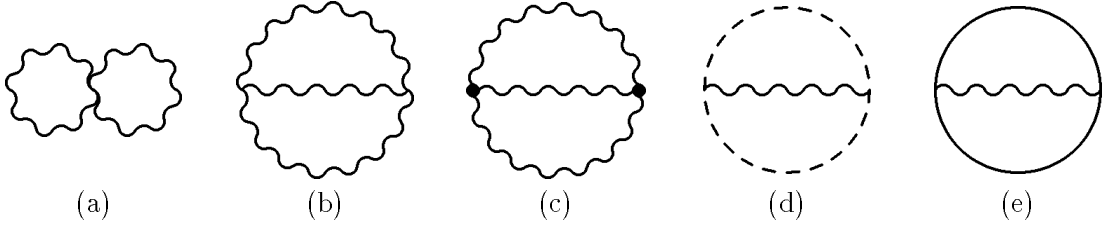}
\end{center}
\caption{Feynman diagrams of 2 Loop corrections to the effective action}
\label{Fig:2LoopCorrections}
\end{figure}

We expand matrices in terms of the
tensor product of matrix spherical harmonics:
\beqa
&&A_{\mu}=f(p_{\mu}+\sum_{jmpq}a^{\mu}_{jmpq}Y_{jm}\otimes Y_{pq}) ,\n
&&\psi=f^{3\over 2}\sum_{jmpq}\psi_{jmpq}Y_{jm}\otimes Y_{pq} ,
\eeqa
where 
\beqa
&&p_{\mu}=j_{\mu}\otimes 1  ~~(\mu=8,9,0),\n
&&p_{\mu}=1\otimes \tilde{j}_{\mu} ~~(\mu=1,2,3),\n
&&other ~~p_{\mu}'s=0.
\eeqa
The summations over $j$ and $p$ run up to
$j=2l_1$ and $p=2l_2$ respectively
where we assume $N=(2l_1+1)\times (2l_2+1)$.
We exclude the singlet state $j=p=0$
in the propagators in this paper.

We adopt the following representation of $Y_{jm}$:
\begin{eqnarray}
&&(Y_{jm})_{ss'}=(-1)^{l-s}
\left(\begin{array}{ccc}
l&j&l_{} \\
-s&m&s'
\end{array}\right)
\sqrt{2j+1} .
\end{eqnarray}
where they are normalized as
\beq
\mbox{Tr}\;Y_{j_1m_1}Y_{j_2m_2}=(-1)^{m_1}\delta_{j_1,j_1}
\delta_{m_1,-m_{2}} .
\eeq
The cubic couplings of the matrix spherical harmonics
can be evaluated as
\beqa
&&Tr[Y_{j_1m_1}Y_{j_2m_2}Y_{j_3m_3}]\n
&&=(-1)^{2l}\sqrt{(2j_1+1)(2j_2+1)(2j_3+1)}\n
&&\times\left(\begin{array}{ccc}
j_1&j_2&j_3\\
m_1&m_2&m_3
\end{array}\right)
\left\{\begin{array}{ccc}
j_1&j_2&j_3\\
l&l&l
\end{array}\right\} .
\label{6jsym}
\eeqa
We refer to \cite{Edm} for $( 3j   )$ and  $\{6j \}$ symbols.

\subsubsection*{bosonic propagators }
From the quadratic terms in the gauge fixed action, we can read propagators
of gauge boson modes $a^{\mu}_{jmpq}$ and ghost modes 
$b_{jmpq}$, $c_{jmpq}$ as follows
\begin{eqnarray}
\langle\; a^{\mu}_{j_1m_1p_1q_1}a^{\nu}_{j_2m_2p_2q_2}\;\rangle&=&
\frac{1}{f^{4}}\;\frac{(-1)^{m_1+q_1}}{j_1(j_1+1)+p_1(p_1+1)}\;
\delta^{\mu\nu}\delta_{j_1j_2}\delta_{p_1p_2}
\delta_{m_1-m_{2}}\delta_{q_1-q_{2}} 
,
\n
\langle\; c_{j_1m_1p_1q_1}b_{j_2m_2p_2q_2}\;\rangle&=&
\frac{1}{f^{4}}\;
\frac{(-1)^{m_1+q_1}}{j_1(j_1+1)+p_1(p_1+1)}\;
\delta_{j_1j_2}\delta_{p_1p_2}\delta_{m_1-m_{2}}\delta_{q_1-q_{2}}
.
\end{eqnarray}
In terms of fields
\begin{eqnarray}
a^{\mu}_{}&=&\sum_{jmpq}\;a^{\mu}_{jmpq}Y_{jm}\otimes Y_{pq}, \n
b_{}&=&\sum_{jmpq}\;b_{jmpq}Y_{jm}\otimes Y_{pq}, \n 
c_{}&=&\sum_{jmpq}\;c_{jmpq}Y_{jm}\otimes Y_{pq} ,  
\end{eqnarray}
propagators become 
\begin{eqnarray}
\langle\; a^{\mu}_{}\;a^{\nu}_{}\;\rangle&=&
\frac{1}{f^{4}}\;
\sum_{jmpq}\;\frac{(-1)^{m+q}}{j(j+1)+p(p+1)}\;
\delta^{\mu\nu}
(Y_{jm}\otimes Y_{pq})(Y_{j-m}\otimes Y_{p-q}) ,
\n
\langle\; c_{}\;b_{}\;\rangle&=&
\frac{1}{f^{4}}\;\sum_{jmpq}\;
\frac{(-1)^{m+q}}{j(j+1)+p(p+1)}\;
(Y_{jm}\otimes Y_{pq})_{}(Y_{j-m}\otimes Y_{p-q})_{} .
\end{eqnarray}

\subsection*{contribution from 4-gauge boson vertex (a)}

The interaction vertex which involves 4 gauge bosons is
\begin{equation}
V_4 = - \frac{f^4}{4} \textrm{Tr} \, [ a_{\mu},a_{\nu} ] [
a_{\mu},a_{\nu} ] .
\end{equation}
It gives rises to the following contribution
\begin{eqnarray}
&&< - \frac{1}{1!} V_4 >_{\textrm{1PI-2loop}} \n
& = &  \frac{1}{4} (10^2-10) 
\frac{1}{f^4}  
\sum_{j_1 j_2 } \sum_{m_1 m_2 } 
\sum_{p_1 p_2 } \sum_{q_1 q_2 }  (-)^{m_1 + m_2+p_1+p_2}\n
&&
\times
\frac{ \textrm{Tr} \, 
[ Y_{j_1 m_1}Y_{p_1 q_1} ,Y_{j_2 m_2}Y_{p_2 q_2} ] 
[ Y_{j_1 -m_1}Y_{p_1 -q_1} ,Y_{j_2 -m_2}Y_{p_2 -q_2} ]}
{(j_1 (j_1+1)+p_1(p_1+1)) (j_2 (j_2 +1)+p_2(p_2+1))} .
\eeqa
Here we can use the completeness condition:
\beq
\sum_{j_3m_3p_3q_3}(-1)^{m_3+q_3}
(Y_{j_3m_3}Y_{p_3q_3})_{ij}(Y_{j_3-m_3}Y_{p_3-q_3})_{kl}
=\delta_{il}\delta_{jk} .
\eeq
We thus find
\beqa
&& - \frac{45}{2}  \frac{1}{f^4} 
\sum_{j_1 j_2 j_3} \sum_{m_1 m_2 m_3}  
\sum_{p_1 p_2 p_3} \sum_{q_1 q_2 q_3} \n
&&
\frac{(\textrm{Tr} \,  
Y_{j_3 m_3} Y_{p_3 q_3} 
 [Y_{j_1 m_1}Y_{p_1 q_1} ,Y_{j_2 m_2}Y_{p_2 q_2}] )^2}
{(j_1 (j_1+1)+p_1(p_1+1)) (j_2 (j_2+1)+p_2(p_2+1))}\n  
& = & - 45
\frac{1}{f^4}
\sum_{j_1 j_2 j_3}\sum_{m_1 m_2 m_3}
\sum_{p_1 p_2 p_3}\sum_{q_1 q_2 q_3}\n
&&
\frac{(2j_1 +1)(2j_2 +1)(2j_3 +1)(2p_1 +1)(2p_2 +1)(2p_3 +1)}
{(j_1 (j_1+1)+p_1(p_1+1)) (j_2 (j_2+1)+p_2(p_2+1))} 
 (1-(-1)^{j_1 +j_2 +j_3+p_1+p_2+p_3})  \n
&&
\times \left(\begin{array}{ccc}
j_1&j_2&j_3\\
m_1&m_2&m_3
\end{array}\right)^2
\left(\begin{array}{ccc}
p_1&p_2&p_3\\
q_1&q_2&q_3
\end{array}\right)^2
\left\{\begin{array}{ccc}
j_1&j_2&j_3\\
l_1&l_1&l_1
\end{array}\right\}^2
\left\{\begin{array}{ccc}
p_1&p_2&p_3\\
l_2&l_2&l_2
\end{array}\right\}^2
\n & = & - {45}\frac{1}{f^4} 
\left( F_1^p(l_1,l_2)-F_1^{np}(l_1,l_2)\right) ,
\end{eqnarray}
where
\beqa
F_1^p\left(l_1,l_2\right)&=&\frac{1}{(2l_1+1)(2l_2+1)}
\sum_{j_1j_2}\sum_{p_1p_2}
\n && 
\frac{(2j_1+1)(2j_2+1)(2p_1+1)(2p_2+1)}
{(j_1(j_1+1)+p_1(p_1+1))(j_2(j_2+1)+p_2(p_2+1))} ,\n
F_1^{np}\left(l_1,l_2\right)&=&\sum_{j_1j_2}\sum_{p_1p_2}
(-1)^{j_1+j_2+p_1+p_2}\n &&
\times \frac{(2j_1+1)(2j_2+1)(2p_1+1)(2p_2+1)}
{(j_1(j_1+1)+p_1(p_1+1))(j_2(j_2+1)+p_2(p_2+1))}
\left\{\begin{array}{ccc}
l_1&l_1&j_2\\
l_1&l_1&j_1
\end{array}\right\}
\left\{\begin{array}{ccc}
l_2&l_2&p_2\\
l_2&l_2&p_1
\end{array}\right\} .\n
\eeqa

Here we have used the property of 3j-symbol:
\begin{eqnarray}
\sum_{m_3}\sum_{m_1m_2}
\left(\begin{array}{ccc}
j_1&j_2&j_3\\
m_1&m_2&m_3
\end{array}\right)^2
=\sum_{m_3}\frac{1}{2j_3+1}=1 ,
\end{eqnarray}
and the property of 6j-symbol:
\begin{eqnarray}
&&\sum_{j_1p_1}(2j_1+1)(2p_1+1)
\left\{\begin{array}{ccc}
j_1&j_2&j_3\\
l_1&l_1&l_1
\end{array}\right\}^2
\left\{\begin{array}{ccc}
p_1&p_2&p_3\\
l_2&l_2&l_2
\end{array}\right\}^2\n
&=&\frac{1}{(2l_1+1)(2l_2+1)}
,\n
&&\sum_{j_1p_1}\;(-1)^{j_1+p_1}(2j_1+1)(2p_1+1)
\left\{\begin{array}{ccc}
j_1&j_2&j_3\\
l_1&l_1&l_1
\end{array}\right\}^2
\left\{\begin{array}{ccc}
p_1&p_2&p_3\\
l_2&l_2&l_2
\end{array}\right\}^2\n
&=&\left\{\begin{array}{ccc}
l_1&l_1&j_2\\
l_1&l_1&j_3
\end{array}\right\}
\left\{\begin{array}{ccc}
l_2&l_2&p_2\\
l_2&l_2&p_3
\end{array}\right\}.
\label{singlet}
\end{eqnarray}

\subsection*{contribution from 3-gauge boson vertices (b)}
Firstly we calculate 2-loop 1PI contribution from 3-gauge boson vertices:
\begin{eqnarray}
V_3=\frac{f^{4}}{2}{\mbox Tr}\;[p_{\mu},a_{\nu}][a_{\mu},a_{\nu}]
-\frac{f^{4}}{2}{\mbox Tr}\;[p_{\nu},a_{\mu}][a_{\mu},a_{\nu}].
\end{eqnarray}
The result can be expressed in a compact form as:
\beq
\langle\;{1\over 2}V_3V_3\;\rangle=(10-1){1\over f^4}
\Big(<<{P^2-P\cdot Q\over P^2Q^2R^2}>>_p
-<<{P^2-P\cdot Q\over P^2Q^2R^2}>>_{np}\Big) ,
\label{avr2}
\eeq
where $P,Q,R$ are defined as
\beqa
P_{\mu}Y_{j_1m_1}Y_{p_1q_1}&\equiv& [p_{\mu},Y_{j_1m_1}Y_{p_1q_1}],\n
Q_{\mu}Y_{j_2m_2}Y_{p_2q_2}&\equiv& [p_{\mu},Y_{j_2m_2}Y_{p_2q_2}],\n
R_{\mu}Y_{j_3m_3}Y_{p_3q_3}&\equiv& [p_{\mu},Y_{j_3m_3}Y_{p_3q_3}].
\eeqa
We have also introduced the following average:
\beqa
<<X>>_p
&=&\sum_{j_i,p_i,m_i,q_i}\Psi_{123}^*X \Psi_{123},\n
<<X>>_{np}
&=&\sum_{j_i,p_i,m_i,q_i}\Psi_{132}^*X \Psi_{123},\n
\Psi_{123}&\equiv &
Tr Y_{j_1m_1}Y_{j_2m_2}Y_{j_3m_3}TrY_{p_1q_1}Y_{p_2q_2}Y_{p_3q_3} .
\label{avr1}
\eeqa

Using the following relation,
\beq
P\cdot Q\Psi_{123}={R^2-P^2-Q^2\over 2}\Psi_{123} ,
\eeq
we evaluate (\ref{avr2}) as
\beqa
&&{27\over  2f^4}
\Big(<<{1\over P^2Q^2}>>_p
-<<{1\over P^2Q^2}>>_{np}\Big)\n
&=&{27\over 2}\cdot\frac{1}{f^{4}}
\sum_{j_1j_2j_3}\sum_{m_1m_2m_3}
\sum_{p_1p_2p_3}\sum_{q_1q_2q_3}
(1-(-1)^{j_1+j_2+j_3+p_1+p_2+p_3})\n
&&
\times \frac{(2j_1+1)(2j_2+1)(2j_3+1)(2p_1+1)(2p_2+1)(2p_3+1)}
{(j_2(j_2+1)+p_2(p_2+1))(j_3(j_3+1)+p_3(p_3+1))}\;\n
&&\times
\left(\begin{array}{ccc}
j_1&j_2&j_3\\
m_1&m_2&m_3
\end{array}\right)^2
\left(\begin{array}{ccc}
p_1&p_2&p_3\\
q_1&q_2&q_3
\end{array}\right)^2
\left\{\begin{array}{ccc}
j_1&j_2&j_3\\
l_1&l_1&l_1
\end{array}\right\}^2
\left\{\begin{array}{ccc}
p_1&p_2&p_3\\
l_2&l_2&l_2
\end{array}\right\}^2\n
&=&\frac{27}{2f^{4}}
\Bigg(F_1^p(l_1,l_2)-F_1^{np}(l_1,l_2) 
+F_5^p(l_1,l_2)-F_5^{np}(l_1,l_2)\Bigg) ,
\label{avr4}
\eeqa
where
\beq
F_5^p(l_1,l_2)=F_5^{np}(l_1,l_2)={1\over (2l_1+1)(2l_2+1)}
\sum_{jp\neq 00}
{(2j+1)(2p+1)\over (j(j+1)+p(p+1))^2}.
\eeq

Here we have used the following property of 6j-symbol:
\begin{eqnarray}
&&\sum_{j_1p_1\neq 00}(2j_1+1)(2p_1+1)
\left\{\begin{array}{ccc}
j_1&j_2&j_3\\
l_1&l_1&l_1
\end{array}\right\}^2
\left\{\begin{array}{ccc}
p_1&p_2&p_3\\
l_2&l_2&l_2
\end{array}\right\}^2\n
&=&\frac{1}{(2l_1+1)(2l_2+1)}-
\frac{\delta_{j2,j3}\delta_{p2,p3}}{(2l_1+1)(2l_2+1)(2j_2+1)(2p_2+1)}
,\n
&&\sum_{j_1p_1\neq 00}\;(-1)^{j_1+p_1}(2j_1+1)(2p_1+1)
\left\{\begin{array}{ccc}
j_1&j_2&j_3\\
l_1&l_1&l_1
\end{array}\right\}^2
\left\{\begin{array}{ccc}
p_1&p_2&p_3\\
l_2&l_2&l_2
\end{array}\right\}^2\n
&=&\left\{\begin{array}{ccc}
l_1&l_1&j_2\\
l_1&l_1&j_3
\end{array}\right\}
\left\{\begin{array}{ccc}
l_2&l_2&p_2\\
l_2&l_2&p_3
\end{array}\right\}
-\frac{\delta_{j2,j3}\delta_{p2,p3}}{(2l_1+1)(2l_2+1)(2j_2+1)(2p_2+1)}.
\label{singlet}
\end{eqnarray}
It is because the singlet state is absent
in the gluon propagator which results in the extra terms on the
right-hand side of (\ref{singlet}). Although they cancel
each other in (\ref{avr4}) for $U(1)$ case, it is not the case for $U(n)$
gauge groups.

\subsection*{ghost contribution (d)}
We calculate the contribution from the ghost-gauge boson vertex:
\begin{eqnarray}
V_{gh}=-f^{4}\mbox{Tr}[p_{\mu}, b][c,a_{\mu}] .
\end{eqnarray}
The result is that:
\beq
{1\over f^4}\Big(<<{P\cdot Q\over P^2Q^2R^2}>>_p
-<<{P\cdot Q\over P^2Q^2R^2}>>_{np}\Big) .
\eeq

\subsection*{contribution from cubic vertices (c)}
The cubic vertex which contains the structure constant of 
$SU(2)\times SU(2)$ is
\begin{equation}
V_{cubic} =  \frac{i}{3} f^3 f_{\mu \nu \rho} \textrm{Tr} \, 
[ a_{\mu},a_{\nu} ] a_{\rho}.
\end{equation}
Their contribution is
\begin{eqnarray}
&&< \frac{1}{2!} V_{cubic} V_{cubic}  >_{\textrm{1PI-2loop}}\n
&=&{8\over f^4}\Big(<<{1\over P^2Q^2R^2}>>_p
-<<{1\over P^2Q^2R^2}>>_{np}\Big)\n
& = &  \frac{8}{f^4} 
(F_3^p(l_1,l_2)-F_3^{np}(l_1,l_2)) ,
\end{eqnarray}

where

\begin{eqnarray}
F_3^p\left(l_1,l_2\right) &=&
\sum_{j_1,j_2,j_3}\sum_{p_1,p_2,p_3}\n
&&
\times     {(2j_1 +1)(2j_2 +1)(2j_3 +1)(2p_1 +1)(2p_2 +1)(2p_3 +1)
                          \over 
(j_1 (j_1+1)+p_1(p_1+1))(j_2 (j_2 +1)+p_2(p_2+1))
(j_3 (j_3 +1)+p_3(p_3+1))}\n
&&\times
     \left\{
 \begin{array}{ccc}
  j_1 &  j_2 & j_3 \\
  l_1   &  l_1 & l_1 
 \end{array}
\right\}^2 
     \left\{
 \begin{array}{ccc}
  p_1 &  p_2 & p_3 \\
  l_2   &  l_2 & l_2 
 \end{array}
\right\}^2 , \n
F_3^{np}\left(l_1,l_2\right) &=&
\sum_{j_1,j_2,j_3}\sum_{p_1,p_2,p_3}
\left(-1\right)^{j_1+j_2+j_3+p_1+p_2+p_3} \n
&&\times
{(2j_1 +1)(2j_2 +1)(2j_3 +1)(2p_1 +1)(2p_2 +1)(2p_3 +1)
                          \over 
(j_1 (j_1+1)+p_1(p_1+1))(j_2 (j_2 +1)+p_2(p_2+1))
(j_3 (j_3 +1)+p_3(p_3+1))}\n
&&\times
     \left\{
 \begin{array}{ccc}
  j_1 &  j_2 & j_3 \\
  l_1   &  l_1 & l_1 
 \end{array}
\right\}^2 
     \left\{
 \begin{array}{ccc}
  p_1 &  p_2 & p_3 \\
  l_2   &  l_2 & l_2 
 \end{array}
\right\}^2 .
\label{F3}
\end{eqnarray}

\section*{fermionic contribution (e)}

The fermion propagator is
\beqa
-{1\over f^4}{1\over \Gamma^{\mu}P_{\mu}}
&=&{1\over f^4}{1\over P^2+{i\over 2}F_{\mu\nu}
\Gamma^{\mu\nu}}\Gamma^{\rho}P_{\rho}\n
&=&{1\over f^4}\left({1\over P^2}\Gamma^{\mu}P_{\mu}
+{i\over 2}({1\over P^2})^2f_{\mu\nu\rho}
\Gamma^{\mu\nu\sigma}P_{\rho}P_{\sigma}+
O(({1\over P^2})^2)\right) .
\eeqa
We have expanded it in powers of $1/P^2$ up to
the first nontrivial order which is sufficient
to calculate the leading contribution to the
two loop effective action.

2-loop (1PI) contribution to the effective action from the fermion-boson 
vertices is
\begin{eqnarray}
&&
{1 \over f^4}\Big(
-64<<{P\cdot Q \over P^2Q^2R^2}>>_p
+64
<<{\bar{P}^2\bar{Q}^2\over
(P^2)^2(Q^2)^2R^2}>>_p
+64<<{\bar{P}\cdot\bar{Q}\tilde{P}\cdot\tilde{Q}\over 
(P^2)^2(Q^2)^2R^2}>>_p \n
&&+64<<{P\cdot Q \over P^2Q^2R^2}>>_{np}
-64
<<{\bar{P}^2\bar{Q}^2\over
(P^2)^2(Q^2)^2R^2}>>_{np}
-64<<{\bar{P}\cdot\bar{Q}\tilde{P}\cdot\tilde{Q}\over 
(P^2)^2(Q^2)^2R^2}>>_{np}
\Big)\n &=&
{16 \over f^4} \left[  
2 F_1^p\left(l_1,l_2\right) - 2 F_1^{np}\left(l_1,l_2\right)
+F_4^p(l_1,l_2)-F_4^{np}(l_1,l_2)\right] .
\end{eqnarray}
where
\beqa
&&F_4^p(l_1,l_2)\n
&=&
\sum_{j_1 j_2 j_3} 
\sum_{p_1 p_2 p_3} \n
&&\times {(2j_1 +1)(2j_2 +1)(2j_3 +1)(2p_1 +1)(2p_2 +1)(2p_3 +1)
\over (j_1
(j_1+1)+p_1(p_1+1))^2( j_2 (j_2+1)+p_2(p_2+1))^2 (j_3(j_3+1)+p_3(p_3+1)) }\n
&&\times   \Big(j_3 (j_3+1)p_3 (p_3 +1)+2j_1(j_1+1)p_1(p_1+1)
-4j_3(j_3+1)p_1(p_1+1)\n
&&+2j_1(j_1+1)p_2(p_2+1) +4j_1(j_1+1)j_2(j_2+1) \Big)\n
&&\times
     \left\{
 \begin{array}{ccc}
  j_1 &  j_2 & j_3 \\
  l_1   &  l_1 & l_1 
 \end{array}
\right\}^2 
     \left\{
 \begin{array}{ccc}
  p_1 &  p_2 & p_3 \\
  l_2   &  l_2 & l_2 
 \end{array}
\right\}^2 ,\n
\n
&&F_4^{np}(l_1,l_2)\n
&=&
\sum_{j_1 j_2 j_3} 
\sum_{p_1 p_2 p_3} (-1)^{j_1+j_2+j_3+p_1+p_2+p_3}\n
&&\times {(2j_1 +1)(2j_2 +1)(2j_3 +1)(2p_1 +1)(2p_2 +1)(2p_3 +1)
\over (j_1 (j_1+1)+p_1(p_1+1))^2( j_2 (j_2+1)+p_2(p_2+1))^2
(j_3(j_3+1)+p_3(p_3+1)) }\n
&&\times   \Big(j_3 (j_3+1)p_3 (p_3 +1)+2j_1(j_1+1)p_1(p_1+1)
-4j_3(j_3+1)p_1(p_1+1)\n
&&+2j_1(j_1+1)p_2(p_2+1) +4j_1(j_1+1)j_2(j_2+1) \Big)\n
&&\times
     \left\{
 \begin{array}{ccc}
  j_1 &  j_2 & j_3 \\
  l_1   &  l_1 & l_1 
 \end{array}
\right\}^2 
     \left\{
 \begin{array}{ccc}
  p_1 &  p_2 & p_3 \\
  l_2   &  l_2 & l_2 
 \end{array}
\right\}^2 .
\eeqa

\subsection*{2-loop effective action}

In this way we find the total 2-loop free energy $F(l_1,l_2)$ of
$U(1)$ NC gauge theory on fuzzy $S^2\times S^2$ as follows
\beqa
-F(l_1,l_2)&=&8{1\over f^4}\Big(F_3^p(l_1,l_2)-F_3^{np}(l_1,l_2)
+2F_4^p(l_1,l_2)-2F_4^{np}(l_1,l_2)\Big) .
\eeqa

For $U(n)$ gauge group, we obtain
\beqa
-F(l_1,l_2)&=&{1\over f^4}\left(n^3 (
8F_3^p(l_1,l_2)+16F_4^p(l_1,l_2)
-45F_5(l_1,l_2))\right.\n
&&\left. +n
(8F_3^{np}(l_1,l_2)+16F_4^{np}(l_1,l_2)
-45F_5(l_1,l_2))\right).
\eeqa
We numerically find that planar contributions dominate over
nonplanar contributions in 4d limit. 
\beqa
F_3^p(l,l)&=&3.24+O(1/l) ,\n
F_4^p(l,l)&=&3.96+O(1/l) ,\n
F_3^{np}(l,l)&=& O(1/l) ,\n
F_4^{np}(l,l)&=& O(1/l) ,\n
F_5^{p}(l,l)&=&F_5^{np}(l,l)= O(log(l)/l^2).
\eeqa
As the planar amplitudes are quadratically
divergent, our findings are in accord with generic expectations
in NC gauge theory.

\section*{Appendix B}
\renewcommand{\theequation}{B.\arabic{equation}}
\setcounter{equation}{0}

We evaluate the two loop effective action 
of IIB matrix model in the 4d limit in this appendix.
The principle difference caused by the introduction 
of the Myers term resides in the gauge boson propagators.
The inverse gauge boson propagator in IIB matrix model
with $S^2\times S^2$ background is
\beq
P^2\delta_{\mu\nu} +2if_{\mu\nu\rho}P^{\rho} .
\eeq
Since the leading contribution to the two loop effective
action comes from the quadratically divergent part, 
we can approximate the propagators as follows:
\beq
\delta_{\mu\nu}{1\over P^2}- 2if_{\mu\nu\rho}P^{\rho}({1\over P^2})^2
+ 4I_{\mu\nu}(P)({1\over P^2})^3 .
\label{prpIIB}
\eeq
$I_{\mu\nu}(P)$ are the following transverse tensor in $S^2\times S^2$:
\beq
I_{\mu\nu}(P)=
\bar{\delta}_{\mu\nu}\bar{P}^2-\bar{P}_{\mu}\bar{P}_{\nu}
+\tilde{\delta}_{\mu\nu}\tilde{P}^2-\tilde{P}_{\mu}\tilde{P}_{\nu} ,
\eeq
where $\bar{P}_{\mu}$ and $\tilde{P}_{\mu}$ denote
the components in the first and second three dimensional sub-spaces.
The modification of the propagator is caused by
a single or double insertions
of $-2if_{\mu\nu\rho}P^{\rho}/P^2$ vertices into the minimal propagator.

Strictly speaking there is a zero mode in the gauge boson propagator
which corresponds to the total angular momentum 2 in the both
$SU(2)$. 
We may postpone to integrate this mode by
considering the Wilsonian effective action.
This problem does not modify the leading $O(N)$ term 
of the effective action since it arises
from the quadratically divergent UV contributions as we shall find.

Since we do not have the Myers term in IIB matrix model,
we should exclude the diagram (c). We also need to
consider the effects caused by the modification of
gauge field propagators in (\ref{prpIIB}).

From diagram (a), we find the following extra planar contributions
\beqa
&&-{72\over f^4}G_1\n
&&-{12\over f^4}G_1+{6\over f^4}G_2 ,
\label{4gauvx}
\eeqa
where 
the first and the second line in (\ref{4gauvx}) is due to the 
insertions of $-2if_{\mu\nu\rho}P^{\rho}/P^2$ vertices into the identical
and different propagators respectively.
We also have introduced the following functions.
\beqa
G_1&=&{1\over N}
\sum_{j_1,p_1,j_2,p_2}(2j_1+1)(2p_1+1)(2j_2+1)(2p_2+1)\n
&&\times\Big({1\over j_1(j_1+1)+p_1(p_1+1)}\Big)^2 
{1\over j_2(j_2+1)+p_2(p_2+1)},\n
G_2&=&<<({1\over P^2})^2({1\over Q^2})^2R^2>>_p .
\eeqa

From (b) and (d), we find
\beqa
&&{1\over f^4}(8G_1'+6F_3
+32H_1-32H_2)\n
&&+{1\over f^4}(-4G_1'+2G_2+8F_3-16H_3+12H_4-4H_5) ,
\label{gaugecn}
\eeqa
where 
the first and the second line in (\ref{gaugecn}) is due to the 
insertions of $-2if_{\mu\nu\rho}P^{\rho}/P^2$ vertices into the identical
and different propagators respectively as before.
\footnote{In this evaluation, the following partial integration
formula has been used.
\beq
\sum_mP_{\mu}Y^{\dagger}_{jm}P_{\nu}Y_{jm}
=-\sum_mY^{\dagger}_{jm}P_{\nu}P_{\mu}Y_{jm} .
\eeq
}

In this expression, we have introduced the following functions:
\beqa
G_1'&=&<<({1\over P^2})^2{1\over Q^2}>>_p,\n
H_1&=&<<P_{\mu}P_{\nu}I_{\mu\nu}(R)({1\over R^2})^3{1\over P^2}
{1\over Q^2}>>_p,\n  
H_2&=&<<P_{\mu}Q_{\nu}I_{\mu\nu}(R)({1\over R^2})^3{1\over P^2}
{1\over Q^2}>>_p,\n
H_3&=&<<Q_{\mu}R_{\nu}I_{\mu\nu}(P)
({1\over P^2})^2{1\over Q^2}({1\over R^2})^2>>_p,\n
H_4&=&<<Q_{\mu}Q_{\nu}I_{\mu\nu}(P)
({1\over P^2})^2({1\over Q^2})^2{1\over R^2}>>_p,\n
H_5&=&<<Q_{\mu}R_{\nu}I_{\mu\nu}(P)
{1\over P^2}({1\over Q^2})^2({1\over R^2})^2>>_p .
\eeqa

We can evaluate them as
\beqa
G_1'&=&
G_1+{1\over N}\sum_{j,p}{(2j+1)(2p+1)\over (j(j+1)+p(p+1))^3} ,\n
G_2&=&
\sum_{j_1,j_2,j_3}\sum_{p_1,p_2,p_3}
     (2j_1 +1)(2j_2 +1)(2j_3 +1)(2p_1 +1)(2p_2 +1)(2p_3 +1)\n
&&\times         {(j_3 (j_3 +1)+p_3(p_3+1))   \over 
(j_1 (j_1+1)+p_1(p_1+1))^2(j_2 (j_2 +1)+p_2(p_2+1))^2
}\n
&&\times
     \left\{
 \begin{array}{ccc}
  j_1 &  j_2 & j_3 \\
  l_1   &  l_1 & l_1 
 \end{array}
\right\}^2 
     \left\{
 \begin{array}{ccc}
  p_1 &  p_2 & p_3 \\
  l_2   &  l_2 & l_2 
 \end{array}
\right\}^2 .
\eeqa

From (e)
\beqa
&&{1\over f^4}(64G_1'-32F_3
+64H_2)\n
&&+{1\over f^4}(32G_1'
-16G_2+64H_3) .
\label{fermcn}
\eeqa
Here the first line in (\ref{fermcn}) is due to the
double insertions of $-2if_{\mu\nu\rho}P^{\rho}/P^2$ vertices
into the gauge boson propagator
and the second line is due to the modifications of the
both gauge boson and fermion propagators.

The total two loop effective action in IIB matrix model
for $S^2\times S^2$ space is
\beqa
-F&=&{1\over f^4}(16G_1-8G_2-18F_3+32H_1+32H_2
+48H_3+12H_4-4H_5+16F_4)\n
&=&{1\over f^4}(16G_1-8G_2-18F_3+32H_3+16F_4) ,
\eeqa
since $H_1+H_2=0, H_3+H_4=0$
and $H_3=H_5$.
\footnote{
We can further show that 
$F={2}F_3/f^4$.}
The only remaining independent function we need to evaluate is $H_4$:
\beqa
H_4
&=&{1\over 2}\sum_{j_1,j_2,j_3}\sum_{p_1,p_2,p_3}
     {(2j_1 +1)(2j_2 +1)(2j_3 +1)(2p_1 +1)(2p_2 +1)(2p_3 +1)\over 
(j_1 (j_1+1)+p_1(p_1+1))^2(j_2 (j_2 +1)+p_2(p_2+1))^2
(j_3 (j_3 +1)+p_3(p_3+1))}\n
&&\times   \Big(   - 2{(j_1 (j_1+1))^2 }
-{(j_3 (j_3+1))^2 }
+2j_1 (j_1+1)j_2 (j_2+1) 
+4{j_1 (j_1+1)j_3(j_3+1)}\Big)\n
&&\times
     \left\{
 \begin{array}{ccc}
  j_1 &  j_2 & j_3 \\
  l_1   &  l_1 & l_1 
 \end{array}
\right\}^2 
     \left\{
 \begin{array}{ccc}
  p_1 &  p_2 & p_3 \\
  l_2   &  l_2 & l_2 
 \end{array}
\right\}^2 .
\eeqa

We have numerically estimated the following functions 
in the large $l$ limit as
\beqa
&&2G_1(l,l)-G_2(l,l)= 1.93 +O(1/l),\n
&&H_4(l,l)= 0.822 +O(1/l),\n
&&F(l,l)={1\over f^4}6.53 +O(1/l).
\eeqa

\end{document}